\documentclass{elsart}
\usepackage{epsfig}

\begin{document}
\runauthor{Ramos, Fernando Manuel}
\begin{frontmatter}
\title{Nonextensive Thermostatistics and the $H$-Theorem Revisited}
\author[INPE]{Fernando M. Ramos\corauthref{Someone}}
\author[INPE]{Reinaldo R. Rosa}
\author[INPE]{Luis A. W. Bambace}
\address[INPE]{Instituto Nacional de Pesquisas Espaciais, INPE \\
               12201-970 S\~ao Jos\'e dos Campos - SP, Brazil}
\corauth[Someone]{Corresponding Author: fernando@lac.inpe.br}

\begin{abstract}

In this paper we present a new derivation of the $H$-theorem and the corresponding 
collisional equilibrium velocity distributions, within the framework of Tsallis' nonextensive 
thermostatistics. Unlike previous works, in our derivation we do not
assume any modification on the functional form of Boltzmann's original "molecular chaos 
hypothesis". Rather, we explicitly introduce into the collision scenario, 
the existence of statistical dependence
between the molecules before the collision has taken place, through a conditional distribution $f(\vec{v}_2|\vec{v}_1)$. 
In this approach, different equilibrium scenarios emerge 
depending on the value of the nonextensive entropic parameter.

\end{abstract}

\end{frontmatter}

Boltzmann's kinetic theory, certainly one of the most important developments
in the history of statistical mechanics, has as one of its key ingredients
the "molecular chaos hypothesis", the celebrated (and somewhat controversial) 
{\it{Stosszahlansatz}}. The irreversible nature of Boltzmann's equation is
directly linked to this hypothesis, since molecules in a dilute gas are 
assumed to be correlated only {\it{after}} the collision has taken place [1].

Recently, Lima et al. [2] (see also [3]) advanced an {\it{ad-hoc}} modification of the 
{\it{Stoss\-zahlansatz}}, in order to derive a proof of the $H$-theorem, and to 
obtain the corresponding collisional equilibrium velocity distribution, 
within the framework of Tsallis' nonextensive thermostatistics [4]. 
In this paper, we show that there is no need of modifying the functional form of
the molecular chaos hypothesis for accomplishing this task. All we need, besides
adopting Tsallis' prescription for the functional form of the local entropy, is to 
explicitly assume that molecules are also correlated {\it{before}} the collision.
In other words, we assume that the joint distribution $f(1,2)$ associated with
two colliding molecules, factorizes as $f(1,2)=f_{1,2}=f(1)f(2|1)=f_1 f_{21}$, where
$f_{21}$ contains {\it{all}} the information regarding the correlations
between 1 and 2, before the impact.  

Let us start considering a spatially homogeneous gas of $N$
hard-sphere particles of mass $m$ and diameter $s$, in the absence
of an external force. The state of a dilute gas is kinetically characterized
by the velocity distribution function $f(\vec{v},t)$.
The quantity $f(\vec{v}_1,\vec{v}_2,t) d^3\vec{v}_1 d^3\vec{v}_2$ gives, at each time $t$, the
number of particles in the volume element $d^3\vec{v}_1$ around
the velocity $\vec{v}_1$ {\it{and}} in the volume element $d^3\vec{v}_2$ around
the velocity $\vec{v}_2$. The rate of change of $f(\vec{v}_1,\vec{v}_2,t) d^3\vec{v}_1 d^3\vec{v}_2$
verifies the generalized Boltzmann equation

\begin{equation}
\label{eq1}
\frac{\partial f_{1,2}}{\partial t} = C_q(f_{1,2})~~,
\end{equation}

where $C_q$ denotes the generalized collisional term. 
We also make the usual assumptions: (i) Only binary collisions occur in
the gas; (ii) $C_q(f)$ is a local function of the slow varying
distribution function; (iii) $C_q(f)$ is consistent with the energy,
momentum, and particle number conservation laws.
Now, following standard lines we define

\begin{equation}
\label{eq2}
C_q(f_{1,2}) = \frac{s^2}{2} \int |\vec{V} \cdot \vec{e}| R_q \, d\omega d^3\vec{v'}_1 d^3\vec{v'}_2~~,
\end{equation}

where the primes indicate post collision quantities, 
$\vec{V} = \vec{v}_2 - \vec{v}_1$ is the relative velocity before collision,
$\vec{e}$ denotes an arbitrary unit vector, and $d \omega$ is an
elementary solid angle such that $s^2 d \omega$ is the area of the
collision cylinder [5,6]. The quantity $R_q(f,f')$ is a difference
of two correlation functions (just before and after
collision). At this point we depart from [2], who originally proposed
that $R_q(f,f') = \exp_q(f_1'^{q-1}\ln_qf_1') + 
f_2'^{q-1}\ln_qf_2') - \exp_q(f_1^{q-1}\ln_qf_1) + f_2^{q-1}\ln_qf_2)$, 
where $\exp_q$ and $\ln_q$ are $q$-exponential and $q$-logarithmic 
functions (see below), and assume that 
$R_q(f,f')$ simply satisfies the {\it{standard}} functional form of 
Boltzmann's molecular chaos hypothesis

\begin{equation}
\label{eq3}
R_q(f,f') = f_1' f_{21}' - f_1 f_{21}~~,
\end{equation}

where again primes refer to the distribution function after
collision. Note that, although the functional form is the same, there is a subtle but profound 
difference between ours and Boltzmann's original approach, in the sense that now velocities of colliding
molecules are {\it{not}} statistically independent before the impact. 

For the joint entropy we adopt Tsallis' expression,

\begin{equation}
\label{eq4}
H^{1,2}_q = -k \int f_{1,2}^q \ln_q f_{1,2} \, d^3\vec{v}_1 d^3\vec{v}_2~~,
\end{equation}

where $\ln_q = \frac{f^{1-q} -1}{1-q}$. Note that 
$H^{1,2}_q$ reduces to the standard Boltzmann extensive measure $H^{1,2} = H^{1} + H^{2}$, for
$q = 1$. Taking the partial time derivative of the
above expression we obtain

\begin{equation}
\label{eq5}
\frac{\partial H^{1,2}_q}{\partial t} = -k \int (q f_{1,2}^{q-1} \ln_q f_{1,2} + 1) 
\frac{\partial f_{1,2}}{\partial t} \, d^3\vec{v}_1 d^3\vec{v}_2~~.
\end{equation}

Inserting the generalized Boltzmann equation (\ref{eq1}) into (\ref{eq5}), and using (\ref{eq2}), 
expression (\ref{eq5}) can be rewritten as a balance equation

\begin{equation}
\label{eq6}
\frac{\partial H^{1,2}_q}{\partial t} = G_q(t)~~,
\end{equation}

where the source term $G_q$ reads

\begin{equation}
\label{eq8}
G_q = - \frac{k s^2}{2} \int | \vec{V} \cdot \vec{e}| (1+ q f_{1,2}^{q-1} \ln_q f_{1,2}) R_q \, 
d \omega d^3\vec{v}_1 d^3\vec{v}_2 d^3\vec{v'}_1 d^3\vec{v'}_2~~.
\end{equation}

In order to rewrite $G_q$ in a more symmetrical form, some
elementary operations must be done in the above expression.
Following standard lines [5], we first notice that
interchanging $\vec{v}_1$ and $\vec{v}_2$ does not affect the value of the
integral. This happens because the magnitude of the relative
velocity vector and the scattering cross section are invariants.
Similarly, we may use time-reversal symmetry. Note that this step requires
the change of sign of $R_q$ (inverse collision). Implementing
these operations and symmetrizing the resulting expression,
one may show that the source term can be written as

\begin{eqnarray}
\label{eq9}
G_q = - \frac{k s^2}{4} \int | \vec{V} \cdot \vec{e}| (q f^{q-1}_{1,2} \ln_q f_{1,2} 
- q f'^{q-1}_{1,2} \ln_q f'_{1,2}) \nonumber \\ 
R_q \, d \omega d^3\vec{v}_1 d^3\vec{v}_2 d^3\vec{v'}_1 d^3\vec{v'}_2~~.
\end{eqnarray}

Noting that $f_{1,2}=f_1 f_{21}$, and applying the transformation $f^{q-1} \ln_q f = \ln_{q^\ast} f$, 
where $q^\ast = 2 - q$, we finally find

\begin{eqnarray}
\label{eq10}
G_q = \frac{k s^2 q}{4} \int | \vec{V} \cdot \vec{e}| (\ln_{q^\ast} (f'_1 f'_{21})
- \ln_{q^\ast} (f_1 f_{21})) 
(f'_1 f'_{21} - f_1 f_{21}) \, \nonumber \\
d \omega d^3\vec{v}_1 d^3\vec{v}_2 d^3\vec{v'}_1 d^3\vec{v'}_2~~.
\end{eqnarray}

For positive values of $q$, the $q$-log function is concave. Thus, the integrand in equation (\ref{eq10}) 
is never negative because $(\ln_{q^\ast} f'_1 f'_{21} - \ln_{q^\ast} f_1 f_{21})$ and $(f'_1 f'_{21} - f_1 f_{21})$
always have the same sign. Therefore, we obtain the $H_q$-theorem

\begin{equation}
\label{eq11}
\frac{\partial H^{1,2}_q}{\partial t} \geq 0~~,
\end{equation}

for $q^\ast > 0$ or, equivalently, $0 < q < 2$. 
This inequality describes the route to equilibrium within the framework of
Tsallis' nonextensive formalism. We remark that, since $f_1$ and $f_2$ are not statistically independent,
we cannot say anything about $H^1_q$ (or $H^2_q$) individually, whose value may even decrease during
the relaxation process.

Now, to obtain the equilibrium
$q$-distribution all we need is to set $G_q = 0$. Since the integrand appearing 
in (\ref{eq10}) cannot be negative, the source term will vanish, for any pair of molecules
obeying the energy, momentum, and particle number conservation laws, if and only if

\begin{equation}
\label{eq12}
f_1' f_{21}' = f_1 f_{21}~~,
\end{equation}
or equivalently

\begin{eqnarray}
\label{eq13}
\ln_q f_1' + \ln_q f_{21}' + (1-q) \ln_q f_1' \ln_q f_{21}' = 
& \ln_q f_1 + \ln_q f_{21} + \nonumber \\
(1-q) \ln_q f_1 \ln_q f_{21}~~.
\end{eqnarray}

Clearly, $\ln_q f_1$ and $\ln_q f_{21} + (1-q) \ln_q f_1 \ln_q f_{21}$ are collision invariants. 
Thus, the marginal equilibrium distribution must be of the form

\begin{equation}
\label{eq14}
\ln_q f_1 = a_0 + \vec{a}_1 \cdot \vec{v}_1 + a_2 \vec{v}_1 \cdot \vec{v}_1~~,
\end{equation}

where $a_0$ and $a_2$ are constants and $\vec{a}_1$ is an arbitrary constant
vector. By introducing the barycentric velocity $\vec{u}$, we
may rewrite (\ref{eq14}) as

\begin{equation}
\label{eq15}
\ln_q f_1 = \alpha - \gamma |\vec{v}_1 - \vec{u}|^2~~,
\end{equation}

with a different set of constants. Now, defining $\Gamma_q = e_q(\alpha)$ and $\beta = \frac{\gamma}{(1+(1-q)\alpha)}$, 
we obtain a $q$-generalized Maxwell's distribution 

\begin{equation}
\label{eq16}
f_0(\vec{v}_1) = \Gamma_q \, e_q(- \beta |\vec{v}_1 - \vec{u}|^2) = \Gamma_q \, [1 - (1-q) \beta |\vec{v}_1 - \vec{u}|^2]^{1/1-q}~~,
\end{equation}

where $e_q(x) = [1+(1-q)x]^{1/1-q}$, and $\Gamma_q$, $\beta$ 
and $\vec{u}$ may be functions of the temperature. This result is identical to that obtained
in Ref. [2], using a very different functional form for the molecular chaos hypothesis.
Naturally, Boltzmann's classical scenario, and its corresponding results, are recovered 
in the limit of $q \rightarrow 1$.

Now, assuming that, by collision invariance, we have in equilibrium $\ln_q f_{21} + (1-q) \ln_q f_1 \ln_q f_{21} 
\sim - \beta (|\vec{v}_2 - \vec{u}|^2)$, the $q$-generalized conditional and joint distributions can be easily obtained

\begin{equation}
\label{eq17}
f_0(\vec{v}_2|\vec{v}_1) = \Lambda_q \, e_q \left(\frac{- \beta |\vec{v}_2 - 
\vec{u}|^2}{1 - (1-q) \beta |\vec{v}_1 - \vec{u}|^2}\right)~~,
\end{equation}

and

\begin{equation}
\label{eq18}
f_0(\vec{v}_1,\vec{v}_2) = \Theta_q \, e_q[- \beta (|\vec{v}_1 - \vec{u}|^2 + |\vec{v}_2 - \vec{u}|^2)]~~,
\end{equation}
with $\Theta_q$ and $\Lambda_q$ being the appropriate normalization factors.

Note that, although $\vec{v}_1$ and $\vec{v}_2$ are, by definition, statistically {\it{dependent}}, is easy to
verify that the covariance $\langle \vec{v}_1,\vec{v}_2 \rangle$ equals zero, regardless the value of $q$. A
natural question arises from this somewhat paradoxycal result: what is therefore the effect induced by nonextensivity
on the classical picture of a dilute gas? 

To answer this question we plotted in Fig. 1 the function
$\chi_q(\vec{v}_1,\vec{v}_2)$, where $ f_{1,2} = \chi_q \, f_1 \, f_2$, for three values of the entropic parameter. 
$\chi_q$ acts as a $q$-generalized, velocity-dependent Enskog factor, reducing or enhancing the
collision frequency at certain velocities.  For $q>1$, high-velocity collisions are favored,
the {\it{rms}} velocity increases, the mean collision time decreases, and the velocity distribution
ceases to be a Gaussian, displaying power-law tails. Strong, intermittent fluctuations on the velocity
and the particle density fields are expected to emerge. This behavior is reminiscent to that observed in
granular gases [7]. For $q<1$, we have the opposite scenario: high-velocity collisions are filtered,
the {\it{rms}} velocity decreases, the mean collision time increases, and the velocity
distribution displays a cut-off, meaning that the gas occupies only a subset of its phase space.
A condensate of cold, motionless particles may eventually appear.
Naturally, for $q=1$, we have $\chi_q = 1$, regardless the velocities, what recovers Boltzmann's original 
molecular chaos hypothesis.

\begin{figure}
\begin{center}
\includegraphics[width=10cm,angle=270]{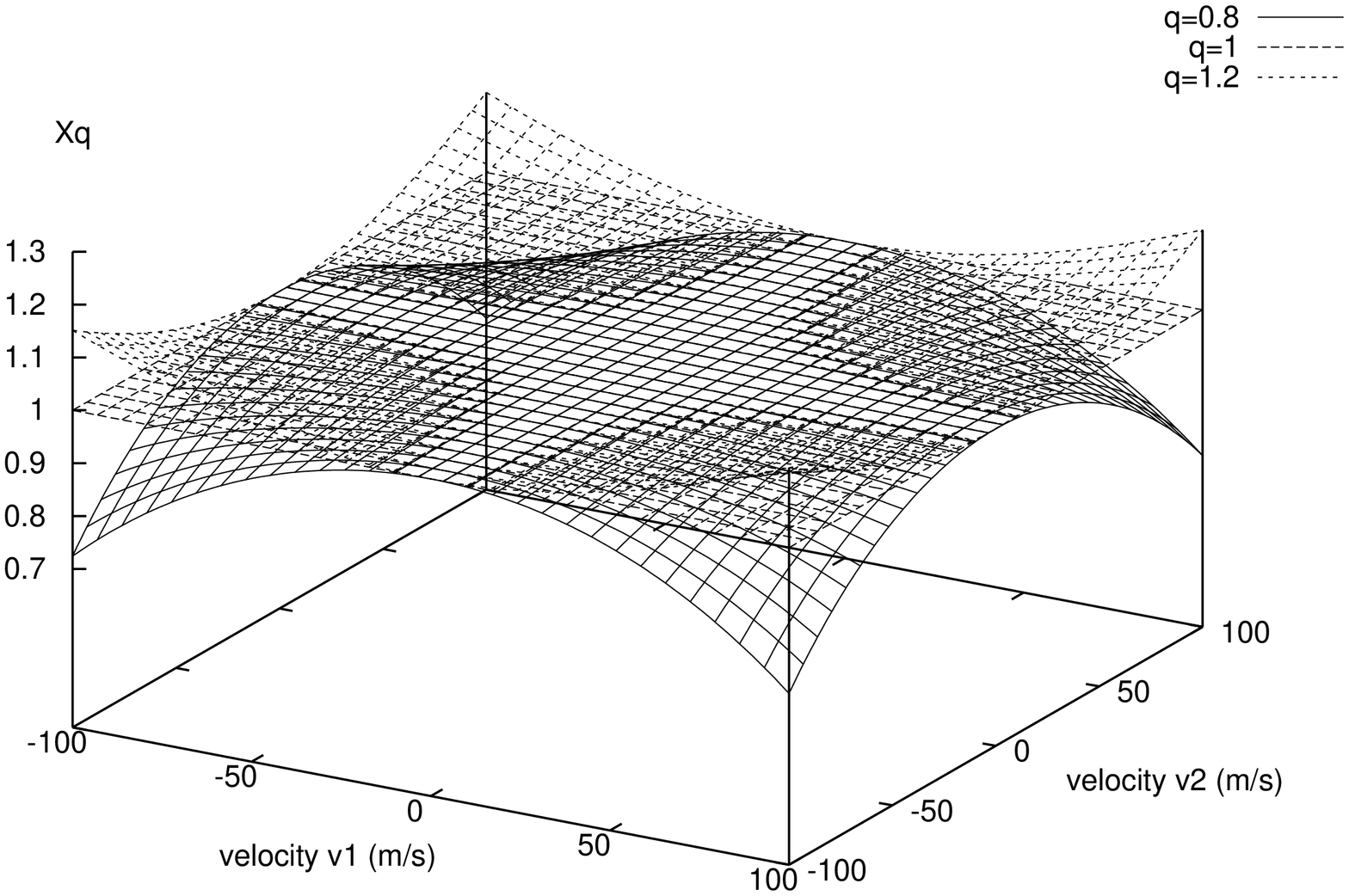}
\caption{Variation of $\chi_q$ for $q=1.2$ (top), $q=1$ (intermediate) and $q=0.8$ (bottom).}
\end{center}
\end{figure}

Summarizing, in this paper we presented a new derivation of the $H$-theorem and the corresponding 
collisional equilibrium velocity distributions, within the framework of Tsallis' nonextensive 
thermostatistics. Unlike previous works, in our derivation we do not
assume any modification on the functional form of Boltzmann's original "molecular chaos 
hypothesis". Rather, we explicitly introduced into the collision scenario, the existence of statistical dependence 
between the molecules before the collision, through a conditional distribution $f(\vec{v}_2|\vec{v}_1)$. 
In this approach, different equilibrium scenarios emerge
depending on the value of the nonextensive parameter.

This work was supported by the Fapesp, CNPQ and CAPES (Brazil). 

\begin{enumerate}

\item[[1]] H. D. Zeh, The Physical Basis of The Direction of Time
(Springer-Verlag, Berlin-Heidelberg, 1992). 

\item[[2]] J. A. S. Lima, R. Silva, A. R. Plastino, Phys. Rev. Lett.
86, 2938 (2001).

\item[[3]] R. Silva, Jr., A. R. Plastino, and J. A. S. Lima, Phys. Lett.
A 249, 401 (1998). 

\item[[4]] C. Tsallis, J. Stat. Phys. 52, 479 (1988). 

\item[[5]] A. Sommerfeld, Thermodynamics and Statistical Mechanics,
Lectures on Theorethical Physics (Academic Press,
New York, 1993), Vol. V. 

\item[[6]] C. J. Thompson, Mathematical Statistical Mechanics
(Princeton University Press, Princeton, New Jersey, 1979). 

\item[[7]] A. Puglisi, V. Loreto, U. Marini Bettolo Marconi, A. Vulpiani,
Phys. Rev. E 59, 5582 (1999).

\end{enumerate}

\end{document}